\documentclass{article}
\usepackage[T1]{fontenc} \usepackage[utf8]{inputenc} \usepackage{ismir,amsmath,cite,url}
\usepackage[normalem]{ulem}
\usepackage{cleveref}
\usepackage{amssymb,mathtools}
\usepackage{graphicx}
\usepackage{color}
\usepackage[export]{adjustbox}
\usepackage[acronym]{glossaries}
\usepackage{multirow}
\usepackage{tabularx}
\usepackage{booktabs}

\usepackage{lineno}

\newacronym{ic}{IC}{information content}

\title{Controlling surprisal in music generation \\ via information content curve matching}

\oneauthor {Mathias Rose Bjare$^1$ \hspace{1.5cm} Stefan Lattner$^2$ \hspace{1.5cm}  Gerhard Widmer $^{1,3}$} {$^1$ Institute of Computational Perception, Johannes Kepler University Linz, Austria\\ $^2$ Sony Computer Science Laboratories (CSL), Paris, France\\ $^3$ LIT AI Lab, Linz Institute of Technology, Austria\\ {\tt 
mathias.bjare@jku.at}}

\def\authorname{M. Bjare, S. Lattner, and G. Widmer}

\begin{document}

\maketitle
\begin{abstract}
In recent years, the quality and public interest in music generation systems have grown, encouraging research into various ways to control these systems. 
We propose a novel method for controlling surprisal in music generation using sequence models.
To achieve this goal, we define a metric called Instantaneous Information Content (IIC). The IIC serves as a proxy function for the perceived musical surprisal (as estimated from a probabilistic model) and can be calculated at any point within a music piece. This enables the comparison of surprisal across different musical content even if the musical events occur in irregular time intervals.
We use beam search to generate musical material whose IIC curve closely approximates a given target IIC.
We experimentally show that the IIC correlates with harmonic and rhythmic complexity and note density. The correlation decreases with the length of the musical context used for estimating the IIC.
Finally, we conduct a qualitative user study to test if human listeners can identify the IIC curves that have been used as targets when generating the respective musical material. We provide code for creating IIC interpolations and IIC visualizations on \url{https://github.com/muthissar/iic}.
\end{abstract}
\section{Introduction}

In music generation, controlling the generation process with user inputs is essential for creating flexible systems that support a creative human/machine co-creation process \cite{aisong}.
Typically, controls are based on low-level features with a direct musical interpretation, for instance, the pitch of a generative synthesizer \cite{nsynth}, or meter, harmony, and instrumentation for symbolic generation \cite{figaro}. 
A high-level musical feature that has received little attention in generative composition systems is musical surprisal — how surprising a musical event is to a listener, given the past musical context. The surprisal tends to be high when the music is complex, when a pitch deviates from the prevailing tonality, or when there is a variation in rhythm \cite{complic,ourssample}. As such, musical surprisal shares similarities with musical complexity, 
however, it is importantly also affected by \textit{learning}: Repeating complex musical content can lead to decreased surprisal on the repetitions as a result of learning \cite{bjaretismir}. 
In contrast, the musical content and the complexity remain unchanged across repetitions.

Studies suggest that the amount of musical surprisal needs to be balanced for music to be deemed preferable \cite{musexp, ucurverhythm}, which is typically achieved by balancing regularity and novelty \cite{creativity}. Being able to control surprisal in generated music might help users create compositions that balance regularity and novelty and thus suit listeners' preferences. In addition, if this can be controlled, rather low surprisal could be used indirectly to induce repetitions in machine-generated music and high surprisal to produce novel parts, possibly with high perceived complexity.

In \cite{meyer}, it was proposed to quantify the surprisal of a musical event by its Information Content (IC) conditioned on past musical events. 
For that, a sequence of musical events is modeled as a stochastic process, where the conditional distribution and, hence, the conditional IC can be estimated.
As such, a surprising event is an event that is unlikely to occur under the estimated distribution given the past musical context. 
In the works of \cite{pitch_per}, the authors find correlations between the IC of a variable-order Markov model (called IDyOM) \cite{idyom_conklin} and perceived surprise in a controlled pitch anticipation experiment. A correlation between high IC and tonal and rhythmic complexity was shown in \cite{complic,ourssample}.

This indicates that the IC of trained sequence models can be used as a proxy for human perception of musical surprisal and that its measurement can identify musical complexity and regularities. This paper proposes a novel framework for generating music with user control over the IC.
Specifically, we define an \textit{Instantaneous Information Content} (IIC) measure, which can be calculated at any time point based on the IC of musical events in the recent past and approximates a causal information density. We use the IIC as a fitness score to direct a beam search toward generating samples following a given IIC target curve. Our sampling strategy can be used with any pretrained auto-regressive generative music model.
We demonstrate our approach in symbolic classical music generation using a pretrained PIA model \cite{pia} and show quantitatively that our approach can generate samples that follow IIC curves extracted from real data. We conducted a qualitative study to test if humans can identify simple IIC curves used for generation. Finally, we analyze relationships between IIC and harmonic, rhythmic, and note density complexity.

\section{Methods}\label{sec:introduction}
In the following, we describe a method for IC-controlled token sequence generation. Let $\text{IC}^{*} \left( t \right)$ be a target curve with support in the time interval $[0,T]$, representing the desired information content over time of a generated sequence of tokens $\mathbf{x} = x_1,x_2,... x_{n} \in \mathcal{X}$, with a duration of $T$ seconds. `Tokens' are not necessarily individual notes or note onsets but can be any token type commonly used in Transformer-based music generation systems (e.g., \cite{musictransformer,pia,remi}). Also, note that we operate on the physical time dimension, not symbolic (score) time measured, e.g., in beats or number of tokens.

Furthermore, let $q$ be a generative sequence model and  $p$ an autoregressive \textit{critic model}, used for estimating the the $i$'th token's conditional token information content
 \begin{equation}
\text{IC}\left(x_{i} | x_{<i} \right)=  -\log p \left(x_{i} \vert x_{<i} \right),
\label{eq:ic}
\end{equation}
where $x_{<i}=x_{1},x_{2},...,x_{i-1}$. In our context, $p$ will be a Transformer model. The proposed method creates new samples using $q$ with an information content that matches the target curve as measured by $p$.
Our method works as follows: Firstly, we define the \textit{Instantaneous Information Content} (IIC) -- a mapping from a (temporally irregular) token sequence and its information content values to a function representing the musical surprisal in the continuous time domain. Secondly, we define an \textit{IC deviation} -- a metric for comparing the similarity between a sequence's IIC curve and the target curve. Finally, we devise a method for generating token sequences with $q$ that minimize the $IC$ deviation.

\subsection{Instantaneous Information Content }
\subsubsection{Temporal Localization of IC Estimates}
To align the information content of musical events, measured on sequence tokens, with the time-domain target IC ($\text{IC}^{*}$), we face a challenge: IC is calculated on sequence elements, while \(\text{IC}^{*}\) pertains to the time domain. Our solution involves assigning each token a temporal position using a mapping function \(f\), effectively ``\textit{temporally localizing}'' or aligning tokens within the musical timeline. 
Note that $f$ can be constructed by analyzing the specific detokenization method associated with \(\mathbf{x}\)'s tokenization that involves turning a sequence of tokens into a time-based music representation like MIDI\footnote{\url{https://midi.org/midi-1-0-detailed-specification}}. In \cref{sec:expmodeldata}, we present an example of such $f$ using the tokenization of \cite{pia}.

Temporal Localization allows us to map \(\text{IC}\) tokenizations to their respective time points in the music. This is crucial, especially for analyzing tokenizations of symbolic music commonly used with Transformers \cite{musictransformer,pia,remi}, where the decoded musical events do not uniformly align in time. Through this approach, \(\text{IC}\) measured on tokens can be directly compared with the time-domain \(\text{IC}^{*}\), facilitating a coherent analysis across different domains of musical representation.

\subsubsection{Interpolation}
\label{sec:int}
Let $f: \mathbb{N} \times \mathcal{X} \rightarrow \mathbb{R}$ be a localization function, mapping the $i$'th token of sequence $\mathbf{x}\in \mathcal{X}$ to the time domain. The IIC at time $t$ in a piece (represented by token sequence $\mathbf{x}$), is a real number computed by a time interpolation of $\mathbf{x}$'s token ICs:
\begin{equation}
    \text{IIC}(t, \mathbf{x}) = \sum_{f\left(i, \mathbf{x} \right) < t}
    \lambda \left(t - f\left(i, \mathbf{x} \right), i \right) \cdot \text{IC}\left(x_i \vert x_{<i} \right).
    \label{eq:ic_int}
\end{equation}
$\lambda \left(t, i \right) $ defines a weighting of the information of the $i$'th token and the constraint $f(i, \mathbf{x})<t$ ensures causality. As a result, the IIC at any time step $t$ is a weighted sum of IC values of past events, using a weighting kernel $\lambda$. 

The choice of the critic model $p$ in combination with the weight function $\lambda$ defines different perceptual models of the instantaneous information content. 
We propose to choose $\lambda$ so that the recent past is weighted higher than the remote past. More specifically, we define $\lambda$ as a window function centered around $t$ and equal to zero at time steps greater than $t$. 
In this initial work, we chose a Hann window for the following reasons: As it is (half) bell-shaped, it is insensitive to inaccuracies in the temporal localization of recent events. It is smooth at the boundaries, preventing sudden drops as events ``leave'' the window.

Using the IIC, we quantify the \textit{segment surprisal} of segment $[t_{1},t_{2}]$ by the $L^{1}$ norm of the IIC with support restricted to $[t_{1},t_{2}]$ by calculating:
\begin{equation}
    \left\lVert \text{IIC}  \mid_{t_1}^{t_2} \right\rVert_{1}  = 
    \int_{t_1}^{t_2} \left\vert \text{IIC}\left(t, \mathbf{x} \right) \right\vert dt.
    \label{eq:ic_segment}
\end{equation}
In \cref{sec:expanaliic}, we compare segment surprisal with segment-based complexity metrics. 

\subsection{IC Deviation}
Given a sample $\mathbf{x}$, the \textit{IC deviation} of $\text{IIC}(\cdot, \mathbf{x})$ from the target $\text{IC}^{*}$ is defined as the $L^{1}$ norm of their function difference:
\begin{equation}
\lVert \text{IC}^{*} - \text{IIC} \rVert_{1}  = 
    \int_{0}^{T} \left\vert \text{IC}^{*} \left( t \right) - \text{IIC}\left(t, \mathbf{x} \right) \right\vert dt.
    \label{eq:ic_dev}
\end{equation}
Which is equal to zero if $\text{IC}^{*} = \text{IIC}(\cdot, \mathbf{x})$ almost everywhere, implying that minimizing \cref{eq:ic_dev}, aligns the target curve $\text{IC}^{*}$ with the $\text{IIC}$ curve. In practice, we compute \cref{eq:ic_dev} by the Riemann sum:
\begin{equation}
    \lVert \text{IC}^{*} - \text{IIC} \rVert_{1}   \approx \sum_{i=1}^{m} \left\vert \text{IC}^{*} \left( t_i \right) - \text{IIC}\left(t_{i}, \mathbf{x} \right) \right\vert \Delta t,
    \label{eq:ic_dev_con}
\end{equation}
where $m\Delta t = T$. 
\subsection{Information Content Conditioned Sampling}
\label{sec:beamsearch}
We can now rank sequences of different lengths according to their proximity to the target $\text{IC}^{*}$ using \cref{eq:ic_dev_con}. We use this to guide a beam search to follow the target curve. The beam search is done in iterations. At each iteration, we generate $k$ continuations of the best-performing sample from the last iteration (initially the empty sequence) in parallel. We stop expanding the continuation when the duration of the newly generated content exceeds a predefined step size $t'$. We then evaluate \cref{eq:ic_dev_con} and keep only the continuation with the lowest IC deviation for the next iteration\footnote{Practically, in beam search iteration $i$, we evaluate the integral of \cref{eq:ic_dev} from $0$ to $it'$.}. We stop when the generation's duration is $T$.

\section{Experiments}
\subsection{Model and Data}
\begin{figure}[t]
    \centering
    \includegraphics[width=.95\linewidth]{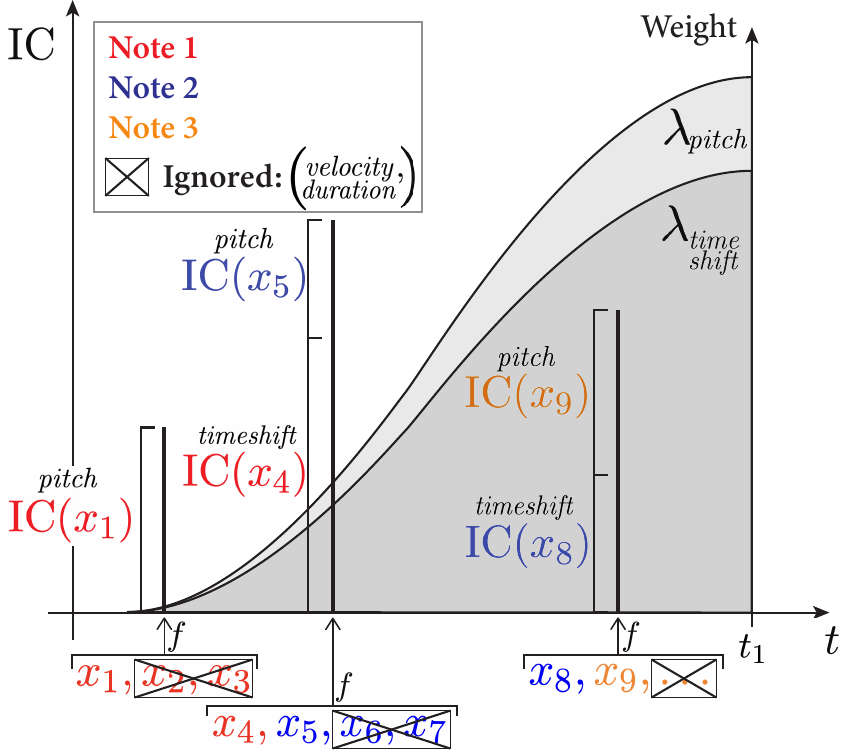}
    \caption{The temporal localization function $f$ and the weight function $\lambda$, involved in computing the $\text{IIC}$ of $x_1,x_2,...$, a sequence of three notes, at time $t_1$.}
    \label{fig:ic_int}
\end{figure}
\label{sec:expmodeldata}
 All experiments are performed with a PIA Transformer model \cite{pia}, a symbolic music generation system pretrained on expressive classical piano performances. 
The model was trained on data consisting of 1,184 MIDI files of expressive music recorded with high precision on a Yamaha Disklavier \cite{maestro}, as well as a larger dataset of 10,855 MIDI files containing automatically transcribed piano performances \cite{giantmidi}. 
For evaluation, we use the dataset of \cite{batik}, consisting of performances of 36 Mozart piano sonata movements. The midi files are tokenized using a \textit{structured MIDI encoding} \cite{pia}, where midi notes, sorted by their onset times, are serialized successively using four tokens $\mathit{Pitch,Velocity,Duration,Timeshift}$ in that order. Therefore, every fourth token represents the same token type. $\mathit{Pitch}$ is an integer describing the 88-note pitch values on the piano. $\mathit{Velocity}$ is an integer describing the 128 possible midi velocity values. $\mathit{Duration}$ is an integer representing quantized note duration in seconds:
$\{0.02, 0.04, ..., 1.0, 1.1, ..., 5.0, 6.0, ...,19.0\}$. $\mathit{Timeshift}$ is an integer encoding the inter-onset intervals (IOI, i.e., the time durations between subsequent note onsets). $\mathit{Timeshift}$ is quantized similarly to the duration token, with the addition of an extra symbol representing a time shift of zero, allowing the model to understand that notes less than $0.02$ seconds apart are to be played concurrently. In contrast to the PIA model described in \cite{pia}, which does non-causal inpainting, we use a causal Transformer based on the Perceiver IO architecture \cite{percieverio} and do continuation generation\footnote{The model generates sequences using an initial context of real music.}. We make these modifications such that the IC calculations ignore future observations. We use the same pretrained model both as the \textit{generator} model $q$ and the \textit{critic} model $p$ and leave the exploration of other critic models for future work. 

\subsection{IIC}
The elements involved in computing the IIC are given in \cref{fig:ic_int}. For IIC calculations, we choose to consider only the surprisal of $\mathit{Pitch}$ and $\mathit{Timeshift}$ tokens, such that the token's IC represents the surprisal of pitches and IOI. We ignore $\mathit{Velocity}$ and $\mathit{Duration}$ tokens because they contribute less to the perception of surprise, being mostly related to the performance dimensions dynamics and articulation. This is achieved in the IIC calculation by setting $\lambda(t,i) = 0$ for $i=2,6,10,...$ and $i=3,7,11,...$ in \cref{eq:ic_int}.  We choose $f$ such that the pitch token contributes to the surprisal function at its note onset time\footnote{The note onset times are found by accumulating the time values associated with previous $\mathit{Timeshift}$ tokens.}, and the timeshift token contributes to its surprisal at the onset of the following note (as an IOI is perceived at the onset of the next note). The remaining weights are then defined by the scaled half Hann window \begin{equation}
    \lambda(t, i) = \begin{cases}c_i \frac{1}{L}cos^{2}(\frac{\pi t}{L}) & \textit{ for $0<t<\frac{L}{2}$, } \\
    0 & \textit{ otherwise}
    \end{cases},
\end{equation} 
where $c_i$ is a weight that takes on two different values for the pitch and timeshift tokens, respectively. $c_i$ is used to weigh the IC of pitches and timeshifts, respectively. For both token types to have equal importance, we estimate a normalization constant empirically by calculating a mean IC over all tokens of the evaluation dataset. The window length is chosen to be $L=4$ so that the weight is zero after 2 seconds.

\subsection{Beam Search Parametrization Study}
Using the beam search strategy described in \cref{sec:beamsearch}, we run initial experiments to determine the effect of parameters associated with the beam search on the similarity between generated samples and target curves $\text{IC}^{*}$ extracted from real music. Specifically, we randomly select 400 snippets of the MIDI files (10 seconds long) and create the IIC curve associated with those snippets. Then, we generate four new samples using our beam search and evaluate the $\text{IC}$ deviation between the IIC curve induced by the real data and the generated data. We discretize the integral in the $\text{IC}$ deviation (see \cref{eq:ic_dev_con}) with $\Delta t = 0.1s$.

To investigate the effect of the step size $t'$, we fix the number of continuations generated in parallel to $k=16$ to reduce computation. To investigate the importance of the number of parallel generated samples $k$, we use a fixed step size of $t'=0.3s$. 

We find that in cases where the generation model $q$ and the critic model $p$ are the same ($p=q$), it is challenging to sample a single continuation $x_{i},x_{i+1},...,x_{i+m}$ (using $q$) that has a high segment surprisal $\lVert \text{IIC} \mid_{f(x_{i})}^{f(x_{i+m})}\rVert_1$ (measured by $p$),
precisely because the probability of sampling such a continuation is low.

To sample low-probability tokens more efficiently, we propose a heuristic that alters the entropy\footnote{Entropy is the expectation of IC.} of the generating distribution $H(q)$ using a temperature parameter dynamically set using the IIC.
Specifically, in iteration $i-1$ of the beam search, we measure $\text{IC}^{*}(it')$, the target IC at the time where the generation of the continuations halts next time, and calculate a target entropy:
\begin{equation}
    H_{\mathit{target}} = \min\left(\frac{\text{IC}^{*}(it')}{C_{H}}, H_{\max}\right),
\end{equation} where $C_{H}$ is a constant parameter to be estimated and $H_{\mathit{max}}$ is the entropy of the uniform distribution. We then fix $q$'s entropy to the target entropy $H_{\mathit{target}}$ by searching for a temperature $r$ such that $H_{\mathit{target}} = H(q) = H(\mathit{softmax}(l/r))$ with binary search, where $l$ are the logits of the neural network. Note that temperature is only used for the generator $q$ and not for the critic model $p$.

\subsection{Qualitative Evaluation}

\label{sec:expuserstudy}
\begin{figure*}
    \centering
    \includegraphics[trim={.2cm 0 0 0},clip,width=\linewidth,frame]{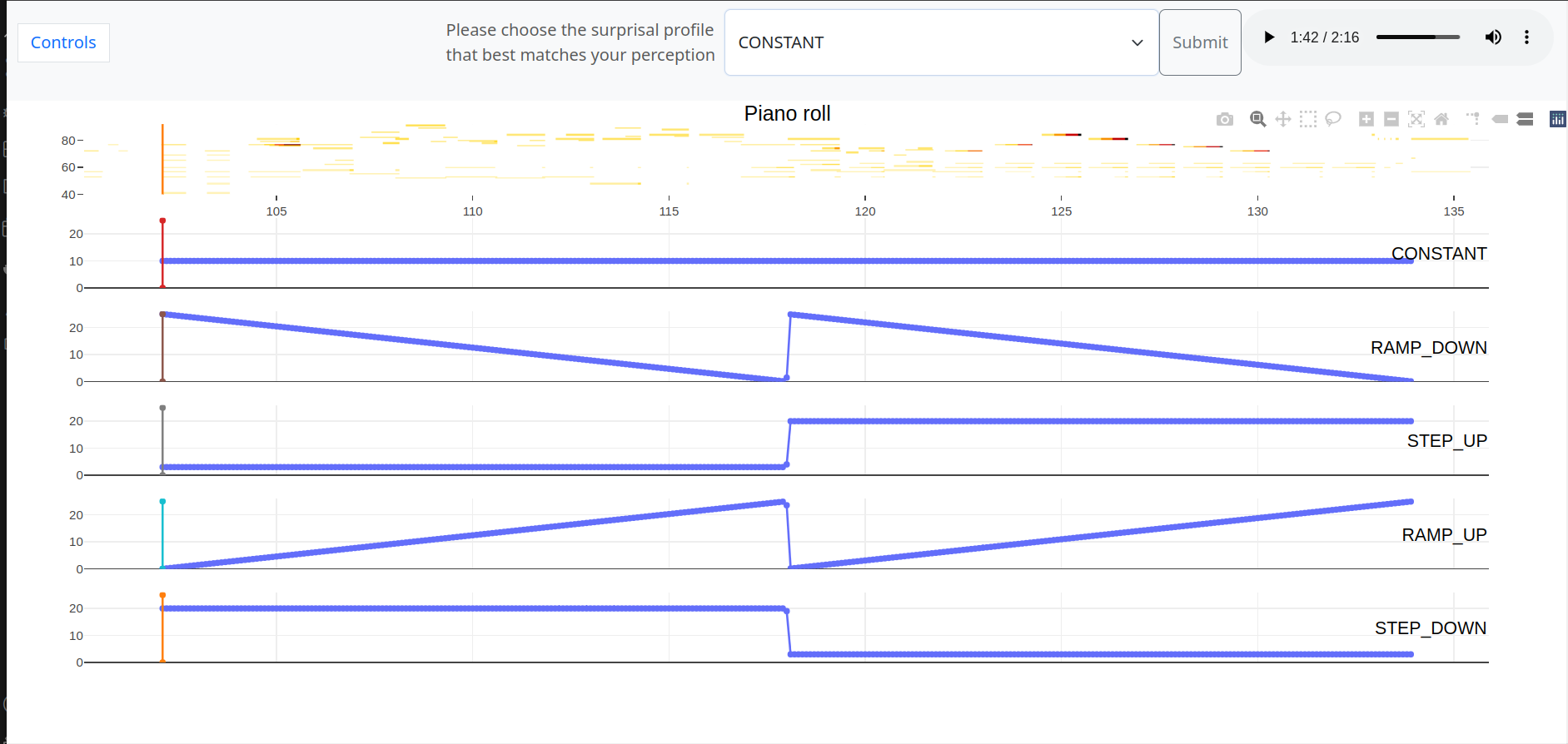}
    \caption{Example page of the user study with a generated musical section and five target curves to choose from.}
    \label{fig:user_study}
\end{figure*}

We conducted an online user study to investigate if the IIC curves computed on generated and real music correspond to users' experience of being musically surprised.

Firstly, we present the participant with a musical section generated by our method using one of five target curves.
The participant is then tasked to select the IIC curve that best describes their perceived surprise when listening to the section. Secondly, we present the user with a segment of real music and IIC curves extracted from real music, one of which corresponds to the music segment. The user's task is to identify the corresponding curve.

The experiment is conducted on a website
that, after an initial experiment description, asks the user for their years of musical training (more or less than five years). Then, it shows an example of a generated piano music section and the surprisal curve used as a target for the generation (together with a textual explanation). 

The participant is then presented with five pages, like the one in \cref{fig:user_study}. Each presents a musical section generated using one of five simple target curves. The participant is asked to identify which of the five curves they think has been used to generate the section. The final page contains a $10$-second segment of real piano music from the evaluation set and two IIC curves, one corresponding to the piano music and the other to a randomly selected 10-second segment from the evaluation dataset.

The samples for the first five pages are generated as follows: As contexts for the model, we select the first 13 measures of Mozart K.331, 1st mvt. and the first 16 measures of K.332, 2nd mvt.~from the evaluation dataset and generate 200 samples for every combination of the two musical contexts and the five IIC curves shown on the page, with $C_H = 50$, $t'=0.3s$ and $k=128$ (i.e., the optimal beam search parameters, as shown in \cref{tab:beamsearch}). For each combination, we then select the 25 samples with the lowest $\text{IC}$ deviation for the user study. For the final page of the user study related to real performances, we select 300 different 10-second segments from the evaluation dataset and compute the IIC curves.
The results of the user study will be presented and discussed in \cref{sec:userstudy}

\subsection{Analysis of IIC}

\label{sec:expanaliic}
As discussed in the introduction, IC and surprisal might be related to aspects of musical complexity, but learning effects may lead to a decrease in surprisal in passages with repeated musical content. To investigate these relationships, we designed an experiment to determine if the IIC correlates with harmonic complexity, as quantified by \textit{tonal tension (cloud diameter)} \cite{tonaltension}, where the IIC is calculated using progressively larger segments of musical context.
Tonal tension is calculated for a segment of music by considering its most dissonant pitch class interval, where an interval dissonance is measured as the distance between the interval pitches embedded in a specific Euclidian space where the position is based on the circle of fifths \cite{helix}. 

We extracted one-second segments centered on the onsets of notes in the evaluation dataset. For $i=1,...,1000$, we then compute the Pearson correlation coefficient between the tonal tension and the segment surprisal (see \cref{eq:ic_segment}) of the first $i$ segments within every performance.

In addition, we investigate complexity in terms of note density, i.e., the number of notes per segment. To do so, we use the same setup as for tonal tension but count the number of notes within one-second segments.

Finally, we investigate rhythmical complexity using the  \textit{IOI histogram entropy} of measures \cite{ioient}. We choose this measure over other structural rhythmical complexity measures \cite{rhythmcompstruc1,rhythmcompstruc2,rhythmcompstruc3,rhythmcompstruc4,rhythmcompstruc5,rhythmcompstruc6} since it does not assume the rhythm to be cyclic. 
We follow the same procedure as mentioned above, 
but instead of selecting fixed-sized segments centered around note-onsets, we select segments of one measure based on the measure annotations \cite{batik}. More specifically, we match the 
notes of the performance with its score notes and extract for each measure: 1) the normalized entropy of the score notes IOI histogram and 2) the segment IIC of the measure normalized with the length of the measure. 
The segment boundaries are estimated by the mean onset time of the first and last note in subsequent measures.

\section{Results}

\subsection{Beam Search Parameter Study}

\begin{table*}[t]
\centering
\begin{tabular}{p{1.5cm}rrrrrrrrrr}
\toprule
$t'$ & 0.1s & 0.2s & 0.3s & 0.4s & 0.5s & 0.6s & 0.7s & 0.8s & 1.0s & 2.0s \\
IC dev. & 3.63 & 2.71 & \textbf{2.61} & 2.72 & 2.69 & 2.93 & 3.03 & 3.11 & 3.31 & 3.90 \\
\cmidrule(r){2-11}
$k$ & 1 & 2 & 4 & 16 & 32 & 64 & 96 & 128 &  & \\
IC dev. & 8.41 & 5.90 & 4.36 & 2.61 & 2.14	 & 1.89 & 1.76 & \textbf{1.69} & & \\
\cmidrule(r){2-11}
$C_H$ & 10 & 20 & 30 & 40 & 50 & 60 & 70 & 80 & 120 & No \\
IC dev. & 8.33 & 4.11 & 2.67 & 2.21 & \textbf{2.15} & \textbf{2.15} & 2.25 & 2.45 & 3.12 & 2.61 \\
\bottomrule
\end{tabular}

\caption{IC deviation between target curves $\text{IC}^{*}$ extracted from real music, and IIC curves from continuations generated with different beam search parameters.}
\label{tab:beamsearch}
\end{table*}
In \cref{tab:beamsearch}, we report the mean $\text{IC}$ deviation of samples generated with different beam search step sizes $t'$, numbers of continuations generated in parallel $k$ and $C_H$, constants used for setting the softmax temperature dynamically. 
Bigger step sizes create longer continuations with high $\text{IC}$ deviation variance, resulting in worse performance. The lowest values ($t'=0.1s,0.2s$) also worsen $\text{IC}$ deviation, likely because sampled notes exceed the timestep, causing inaccuracies in the next beam search iteration.
For the number of continuations generated in parallel $k$, we find that the $\text{IC}$ deviation always decreases with higher $k$. This is not surprising as the model has more candidate continuations to choose from. The decrease flattens out as seen by the small $\text{IC}$ deviation differences when $k\geq64$. For the dynamic temperature, we find that $C_H=50,60$ reduces the $\text{IC}$ deviation compared to using no temperature scaling (marked with "No" in \cref{tab:beamsearch}).

\subsection{Qualitative Results}
\label{sec:userstudy}
\begin{table*}
    \centering
    \begin{tabular}{lrrrrrrr}
\toprule
     \multicolumn{6}{c}{IIC Curves} & \\
 & CONSTANT & RAMP\_DOWN & STEP\_UP & RAMP\_UP & STEP\_DOWN & Gen. all curves & Real \\
 \cmidrule(r){2-6}
 \cmidrule(r){7-8}
$F_1$ & 0.53 & 0.41 & 0.71 & 0.48 & 0.49 & 0.52 & 0.71 \\
\#True & 36 & 34 & 29 & 29 & 24 & 152 & 21 \\
\#Pred & 36 & 33 & 31 & 30 & 22 & 152 & 21 \\
\bottomrule
\end{tabular}

    \caption{Results from the user study reported as $F_{1}$-score of identifying the: IIC curve used for generation, the IIC curve of real music.  
    }
    \label{tab:user_study}
\end{table*}
The user study results reported as a binary classification of finding the correct curve, among the curves described in \cref{sec:expuserstudy}, are presented in Table \ref{tab:user_study}. $29$ users participated, $23$ participants had more than $5$ years of musical training, and $6$ participants had less than $5$ years of experience. $152$ generated samples and $21$ samples of real music were classified in total. 
Due to the imbalance in the number of untrained and trained participants and since we found little difference in the classification performance between the groups, we combined their results in the table.

\begin{figure}
    \centering
    \includegraphics[width=.8\linewidth]{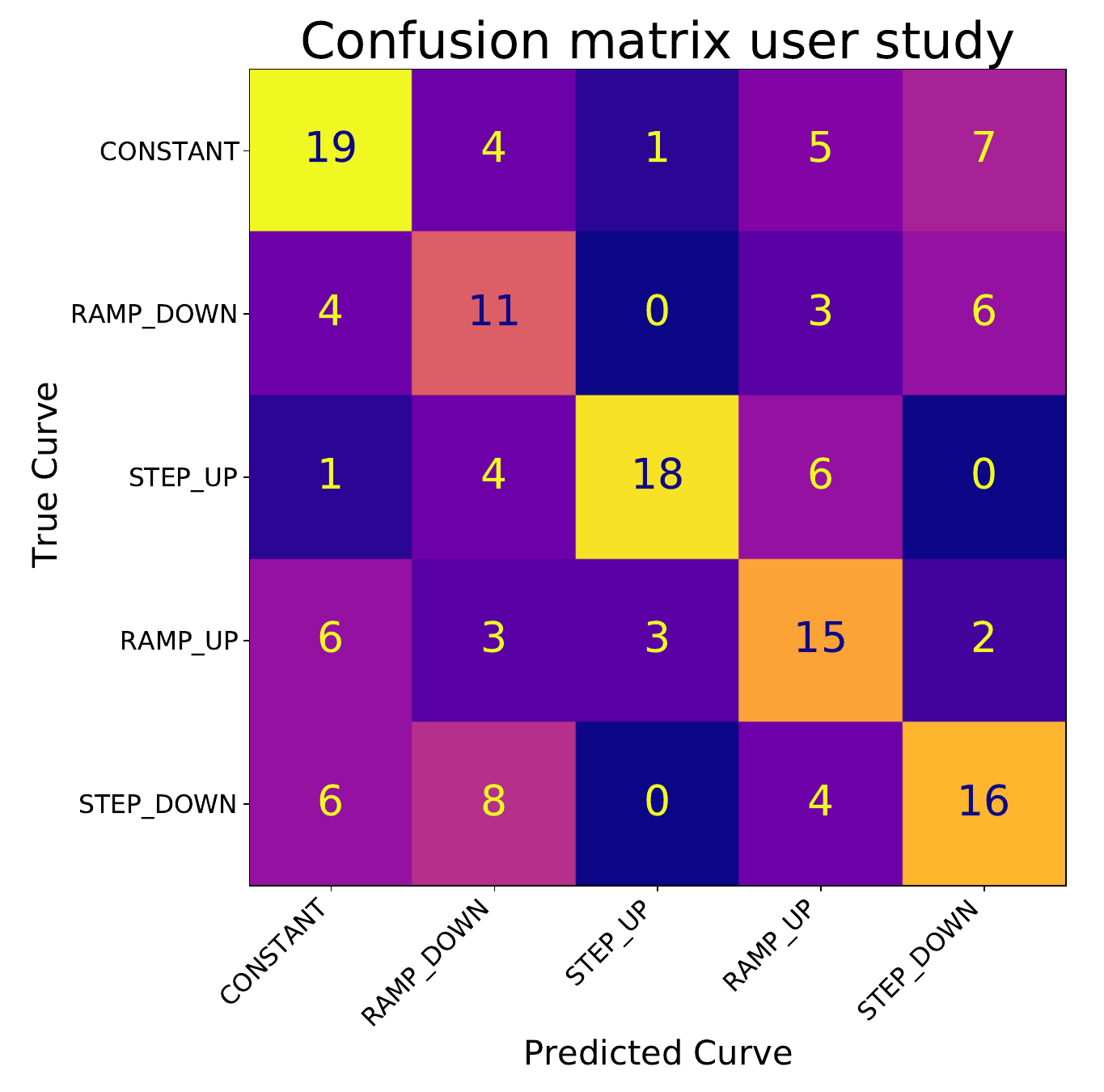}
    \caption{The confusion matrix for users identifying the $\text{IC}^{*}$ curves used to generate the music examples. 
    }
    \label{fig:user_study_confusion}
\end{figure}

The overall $F_1$-score was reported as $0.52$ for generated data and $0.71$ for real data, which is reasonably above the proportions $0.2$ and $0.5$, being the $F_1$ scores of random classifiers, with 5 and 2 classes respectively. The results for the individual curves show difficulty differences in classifying the different curve types, with RAMP\_DOWN having the lowest $F_1$-score of $0.41$ and STEP\_UP having the highest $F_1$-score of $0.71$. We therefore investigate the confusion of curves in Figure \ref{fig:user_study_confusion}. We find that the confusion of CONSTANT is evenly distributed on all curves, except for STEP\_UP, which is reasonable since CONSTANT does not share any characteristics with the other curves. 
We furthermore find that generations that start with the same IIC value, either high or low, are confused. This is seen by the 
confusion of RAMP\_DOWN with STEP\_DOWN and the confusion of STEP\_UP and RAMP\_UP. 

\subsection{Analysis of IIC}

\begin{figure}
    \centering
    \includegraphics[width=\linewidth]{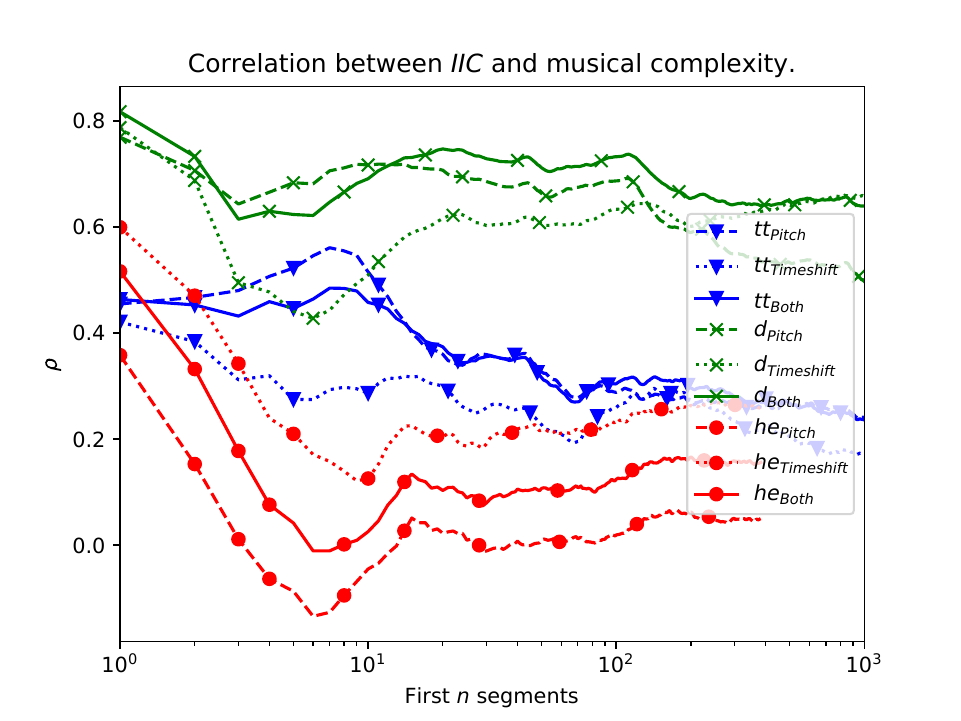}
    \caption{Correlation between IIC and tonal tension  $\mathit{tt}$, note density $d$, and IOI histogram entropy ($he$).         }
    \label{fig:iicanalysis}
\end{figure}

The correlations between IIC and the tonal tension $tt$, note density $d$, and the IOI histogram entropy $he$ were calculated on the first $n$ segments of the 36 evaluation data performances as described in \cref{sec:expanaliic} and reported in \cref{fig:iicanalysis}. We report the results for IIC calculated using $\mathit{Pitch}$ only, $\mathit{Timeshift}$ only, or both token types. 
For $tt$, $d$, and $he_{\mathit{Timeshift}}$, the correlations reported were found significant using a significance level of $0.05$, whereas for $he_{\mathit{Pitch}}$ and $he_{\mathit{Both}}$, the correlations are not significant.

The results show a moderate to high correlation of IIC with all metrics at the beginning of the performances (when $n$ is small). However, these correlations decrease in later parts of the performances (when $n$ is high), likely due to ``learning'' (simulated by longer context) over time.

The highest correlations are found for note density $d$.  This may be explained by the definition of IIC (see \cref{eq:ic_int}) as a weighted sum of token ICs since more tokens per segment simply lead to higher sums.

Considering the different token type combinations, we find that $tt$ is most correlated with IIC calculated using only $\mathit{Pitch}$ tokens and $he$ using only $\mathit{Timeshift}$ tokens. This is reasonable, considering that very dissonant segments and very complex rhythms tend to be associated with $\mathit{Pitch}$ and $\mathit{Timeshift}$ tokens, respectively, which are infrequent in the training dataset, resulting in a high token $\text{IC}$. Interestingly, $tt$ is also correlated with IIC calculated using only $\mathit{Timeshift}$ tokens (encoding IOIs), which might
stem from the critic model facing greater uncertainty in predicting any token type when confronted with a highly harmonic complex context that is infrequent in the dataset.

The curves of $\mathit{Both}$ and $\mathit{Pitch}$ follow each other closely for total tension and note density, indicating that using both $\mathit{Timeshift}$ and $\mathit{Pitch}$ tokens does not significantly reduce the complexity correlations. For rhythmical complexity, using $\mathit{Both}$ tokens instead of $\mathit{Timeshift}$ tokens alone decreases the correlation more.

\section{Conclusion}

In this study, we introduced a novel framework for controlling musical surprisal through Instantaneous Information Content (IIC), which maps token-based surprisal to a continuous time-domain function. Using a beam search algorithm, we demonstrated that our approach can generate music that closely follows predefined IIC curves, effectively aligning generated and target surprisal curves.

Our user study confirmed that participants could reasonably identify target IIC curves from generated music, indicating that our method captures perceptible aspects of musical surprise. Furthermore, our analysis showed that IIC correlates with measures of musical complexity such as tonal tension and note density.

Future work will explore alternative critic models, like personalized models, trained on music that is familiar to the user or models with smaller context windows to more directly control local musical complexity.

\section{Acknowledgments}
The work leading to these results was conducted in a collaboration between JKU and Sony Computer Science Laboratories Paris under a research agreement. The first and third author also acknowledge support by the European Research Council (ERC) under the European Union’s Horizon 2020 research and innovation programme, grant agreement 101019375 (\textit{“Whither Music?”}).

\bibliography{rhythm}

\end{document}